\documentclass[twocolumn,article]{revtex4}%
\usepackage{amsfonts}
\usepackage{amsmath}
\usepackage{amssymb}
\usepackage{graphicx}%
\providecommand{\U}[1]{\protect\rule{.1in}{.1in}}
\begin{document}
\title{Self-organization of dissipationless solitons in negative refractive index
materials }
\author{V. Skarka, N. B. Aleksic, and V. I. Berezhiani}
\affiliation{Laboratory POMA, UMR 6136 CNRS, University Angers, 2, boulevard Lavoisier,
49045 Angers, France}
\affiliation{Institute of Physics, Pregrevica 118, 11000 Belgrade, Serbia }
\affiliation{Andronikashvili Institute of Physics, 6 Tamarashvili, Tbilisi 0177, Georgia}


\begin{abstract}
General nonlinear and nonparaxial dissipative complex Helmholtz equations for
magnetic and electric fields propagating in negative refractive index
materials (NIMs) are derived \textit{ab initio} from Maxwell\textbf{
}equations. In order to describe nonconservative soliton dynamics in NIMs,
such coupled equations are reduced into generalized Ginzburg-Landau equation.
Cross-compensation between the excess of saturating nonlinearity, losses, and
gain renders these self-organized solitons dissipationless and exceptionally
robust. The presence of such solitons makes NIMs effectively dissipationless.

\end{abstract}
\startpage{1}
\endpage{1}
\maketitle

Metamaterials are novel artificial composite structures manufactured in order
to display various peculiar very promising properties \cite{Pendry1}. The most
studied nowadays are negative-refractive-index materials (NIMs) assembled in
such a way to exhibit simultaneously negative effective permittivity
$\varepsilon$ and permeability\ $\mu$ \cite{Veselago},\cite{cloaking}. A
number of challenging optical NIM devices is supposed to work in interaction
with lasers \cite{Optical}. As a consequence, nonlinear effects during the
propagation of electromagnetic (EM) waves in NIMs have to be taken into
account. In particular, in such materials conservative localized solitonic
structures of Schr\"{o}dinger type are considered \cite{Lazar},\cite{Shukla}%
,\cite{Nadia} \cite{Wen}. However, in real media dissipation is always
present. Whenever the absorption is weak, it can be ignored provided that EM
field frequency lays beyond the resonance bands of medium. However, in NIMs
the problem of EM field absorption is the central issue due to the resonant
character of interactions. Persistence of losses prevents also the generation
of Schr\"{o}dinger type solitons in ordinary materials, a fortiori in NIMs.
Numerous highly desirable properties of NIMs like the superresolution
\cite{Pendry1} and the cloaking \cite{cloaking}, are either altered or
annihilated by the dissipation \cite{Webb}. A wide class of dissipative
systems, ranging from nonlinear optics, plasma physics, and fluid dynamics to
superfluidity, superconductivity, and Bose-Einstein condensates,\ can be
modeled by complex Ginzburg-Landau equation \cite{AhmKivs}.

In order to study propagation of EM field in optical NIMs, we derive, in\ this
letter, \textit{ab initio} from Maxwell\textbf{ }equations (ME) general
nonlinear and nonparaxial dissipative complex Helmholtz equations for magnetic
and electric fields. A very promising class of self-organized localized EM
structures are spatiotemporal solitons rendered dissipationless due to the
cross-compensation of the excess of saturating nonlinearity, losses, and gain
\cite{Skarka}. Obtained coupled equations are reduced into generalized
Ginzburg-Landau equation\ in order to describe for the first time
dissipationless soliton dynamics in NIMs. Such a NIM in presence of
dissipationless solitons may be considered as an effectively dissipationless
novel active composite metamaterial.

The propagation of EM radiation in a nonlinear media is described by ME
$\mathbf{rotE=-}\partial\mathbf{B}/\partial t$, $\mathbf{rotH=}\partial
\mathbf{D/\partial}t$ and constitutive relations between the magnetic
induction $\mathbf{B}$ and the magnetic field $\mathbf{H}$ as well as between
the electric field $\mathbf{E}$ and the electric induction $\mathbf{D}$. The
response of the medium to a quasi-monochromatic EM wave is considered. Real
vectorial fields $\mathbf{F=(E,B,H,D)}$ read $\mathbf{F}(\mathbf{r}%
,t)=\mathbf{F}\exp\left(  i\mathbf{k}_{r}\mathbf{r}-\omega t\right)  +c.c.$,
where $\mathbf{F}$ is slowly varying function in space and time and $\omega$
is the carrying pulse frequency. The complex wave vector $\mathbf{k=k}%
_{r}+i\mathbf{k}_{i}$ is determined by the linear dispersion relation
$\mathbf{k}^{2}c^{2}/\omega^{2}=\varepsilon\left(  \omega\right)  \mu\left(
\omega\right)  $. Respectively $\varepsilon$ and $\mu$ are complex
permittivity and permeability of a dissipative medium. \ In NIMs the
refractive index is negative $n=-(\Re\lbrack\varepsilon\mu])^{1/2}%
=k_{r}c/\omega$, hence, the real part of wave vector $k_{r}$ is negative too.
In the case of temporal dispersive media $\varepsilon\left(  \omega\right)  $
and $\mu\left(  \omega\right)  $ are expanded in the series around the
carrying frequency up to the second order \cite{Wen},\cite{Scalora}
$\partial\mathbf{D/}\partial t\approx\varepsilon_{o}(\alpha_{\varepsilon
}\partial\mathbf{E/}\partial t+i\alpha_{\varepsilon}^{\prime}/2\partial
^{2}\mathbf{E/}\partial t^{2}-i\omega\varepsilon\mathbf{E}-i\omega
\mathbf{\varepsilon}_{nl}\mathbf{E)}e^{-i\omega t}$ and $\partial
\mathbf{B/}\partial t\approx\mu_{o}(\alpha_{\mu}\partial\mathbf{H/}\partial
t+i\alpha_{\mu}^{\prime}/2\partial^{2}\mathbf{H/}\partial t^{2}-i\omega
\mu\mathbf{H}-i\omega\mathbf{\mu}_{nl}\mathbf{H)}e^{-i\omega t}$, where
$\alpha_{\varepsilon}=\partial(\omega\varepsilon)/\partial\omega,$
$\alpha_{\varepsilon}^{\prime}=\partial\alpha_{\varepsilon}/\partial\omega,$
$\alpha_{\mu}=\partial(\omega\mu)/\partial\omega,$ and $\alpha_{\mu}^{\prime
}=\partial\alpha_{\mu}/\partial\omega$. In order to take into account the
nonlinear loss and gain\ the nonlinear permittivity $\mathbf{\varepsilon}%
_{nl}\left(  |\mathbf{E}|^{2}\right)  =\Re\lbrack\mathbf{\varepsilon}_{nl}]+$
$i\Im\lbrack\mathbf{\varepsilon}_{nl}]$ and permeability $\mathbf{\mu}%
_{nl}\left(  |\mathbf{H}|^{2}\right)  =\Re\lbrack\mathbf{\mu}_{nl}%
]+i\Im\lbrack\mathbf{\mu}_{nl}]$ are considered as complex. Taking
$\mathbf{\operatorname{curl}}$ of ME and neglecting vectorial terms
($\nabla\left(  \nabla\cdot\mathbf{E}\right)  =0$) yield following equations
rewritten in new variables $\zeta=z$ and $\tau=t-z/v_{g}$ with real group
velocity $v_{g}=\left(  \partial\left(  n\omega\right)  /\partial
\omega\right)  ^{-1}=\left(  \Re\lbrack\omega\left(  \mu\alpha_{\varepsilon
}+\varepsilon\alpha_{\mu}\right)  (2kc)^{-1}]\right)  ^{-1}$,%
\begin{equation}%
\begin{array}
[c]{c}%
2ik\frac{\partial\mathbf{E}}{\partial\zeta}-\frac{W}{c^{2}}\frac{\partial
^{2}\mathbf{E}}{\partial\tau^{2}}+\Delta\mathbf{E}+\frac{\omega^{2}}{c^{2}}%
\mu\mathbf{\varepsilon}_{nl}\mathbf{E}\\
\\
-\omega\mu_{o}\mathbf{\mu}_{nl}\mathbf{k}\times\mathbf{H}+i\frac{\omega^{2}%
}{c^{2}}\Im\lbrack\varepsilon\mu]\mathbf{E}=0,
\end{array}
\end{equation}
and%
\begin{equation}%
\begin{array}
[c]{c}%
2ik\frac{\partial\mathbf{H}}{\partial\zeta}-\frac{W}{c^{2}}\frac{\partial
^{2}\mathbf{H}}{\partial\tau^{2}}+\Delta\mathbf{H}+\frac{\omega^{2}}{c^{2}%
}\varepsilon\mathbf{\mu}_{nl}\mathbf{H}\\
\\
+\omega\varepsilon_{o}\mathbf{\varepsilon}_{nl}\mathbf{k}\times\mathbf{E}%
+i\frac{\omega^{2}}{c^{2}}\Im\lbrack\varepsilon\mu]\mathbf{H}=0,
\end{array}
\end{equation}
with the complex function $W=c^{2}V_{g}^{-2}-0.5\omega\mu\alpha_{\varepsilon
}^{\prime}-0.5\omega\varepsilon\alpha_{\mu}^{\prime}-\alpha_{\mu}%
\alpha_{\varepsilon}$. The second order $z$ derivatives in $\Delta\mathbf{E}$
and $\Delta\mathbf{H}$ should be kept in Eqs. (1-2) not only to account for
nonparaxial pulse dynamics but also to have an adequate description of the
strong dissipation. In best of our knowledge it is for the first time that
such generalized coupled complex Helmholtz equations describing simultaneously
nonparaxial and dissipative effects in NIMs are ab initio derived from ME.
However, investigation of the EM filed dynamics based on the generalized
Helmholtz equations is beyond of intended scope of the current letter and will
be presented elsewhere. Therefore, in what follows only weak dissipation is
considered. As a consequence, the imaginary part of linear permittivity
$\varepsilon_{i}$, permeability $\mu_{i}$, and wave vector $k_{i}$ are small.
Using Drude model of free electron collisions ($\nu_{\varepsilon}$ and
$\nu_{\mu}$), permittivity and permeability read respectively $\varepsilon
=1-\omega_{p}^{2}[\omega(\omega+i\nu_{\varepsilon})]^{-1}$ and $\mu
=1-\omega_{m}^{2}[\omega(\omega+i\nu_{\mu})]^{-1}$ where $\omega_{p}$ and
$\omega_{m}$ are electric and magnetic plasma frequencies \cite{Kivshar}. For
linearly polarized EM pulses ($\mathbf{E}=\widehat{\mathbf{x}}E,\mathbf{H}%
=\widehat{\mathbf{y}}H$) propagating along the $z$ axis ($\mathbf{k}%
\Vert\widehat{\mathbf{z}}$) Eqs. (1-2) become scalar ones. Applying paraxial
approximation ($k>>\nabla$) and neglecting higher order $z$\ derivatives Eqs.
(1-2) can be reduced to generalized Ginzburg-Landau equation in NIMs\ (GLENIM)
for electric field
\begin{equation}%
\begin{array}
[c]{c}%
i\frac{\partial E}{\partial z}-\frac{k_{r}^{\prime\prime}}{2}\frac
{\partial^{2}E}{\partial t^{2}}+\frac{1}{2k}\Delta_{\perp}E+i\varpi\frac
{\Im\lbrack\varepsilon\mu]}{2n}E\\
\\
-i\frac{\Im\lbrack W]}{2n}\frac{\partial^{2}E}{\partial t^{2}}+\frac{\varpi
}{2}\left(  Z\varepsilon_{nl}+\frac{\mu_{nl}}{Z}\right)  E=0
\end{array}
\end{equation}
and the equivalent magnetic field GLENIM obtained using medium impedance
$Z=(\mu/\varepsilon)^{1/2}=E/H$. The following renormalisation is used:
$\omega_{p}\tau\rightarrow t$, $\left(  \omega_{p}/c\right)  \zeta\rightarrow
z$, $\omega/\omega_{p}\rightarrow\varpi$, and $\left(  c/\omega_{p}\right)
^{2}\Delta_{\perp}\rightarrow\Delta_{\perp}=\partial^{2}E/\partial
x^{2}+\partial^{2}E/\partial y^{2}$. The renormalized frequency $\varpi$ runs
in Fig.1 till $\omega_{m}/\omega_{p}=0.8$ \cite{Scalora}. In this range the
group velocity $v_{g}$ is always positive and the phase velocity is negative.
The group velocity dispersion (GVD) $k_{r}^{\prime\prime}=-(n\varpi)^{-1}%
\Re\lbrack W]$ is anomalous, thus, negative till $\varpi=0.7$, changing sign
after (see also Fig.2). As many materials, NIMs also can exhibit cubic Kerr
nonlinearity which may be saturated \cite{Kivshar},\cite{Maluckov}. In order
to prevent pulse collapse the cubic nonlinearity is usually saturated by a
quintic one with the opposite sign $\mathbf{\varepsilon}_{nl}\left(
|E|^{2}\right)  =(\mathbf{\varepsilon}_{r}^{(3)}+i\mathbf{\varepsilon}%
_{i}^{(3)})|E|^{2}-(\mathbf{\varepsilon}_{r}^{(5)}+i\mathbf{\varepsilon}%
_{i}^{(5)})|E|^{4}$ and $\mathbf{\mu}_{nl}\left(  |H|^{2}\right)
=(\mathbf{\mu}_{r}^{(3)}+i\mathbf{\mu}_{i}^{(3)})|H|^{2}-(\mathbf{\mu}%
_{r}^{(5)}+i\mathbf{\mu}_{i}^{(5)})|H|^{4}$. The sign of the real part of
these indices determines the focusing/defocusing properties of the medium
while imaginary parts are related to the nonlinear loss and gain. Nonlinear
permittivity and permeability in synergy define the cubic susceptibility
$\mathbf{\chi}^{(3)}=\mathbf{\varepsilon}^{(3)}Z+\mathbf{\mu}^{(3)}/Z^{3}$ as
well as the quintic one $\mathbf{\chi}^{(5)}=\mathbf{\varepsilon}%
^{(5)}Z+\mathbf{\mu}^{(5)}/Z^{5}$. Corresponding relations for magnetic
susceptibility $\mathbf{\chi}_{H}$ can be obtained using the impedance $Z$.
Therefore, the system of coupled Eqs. (1-2) is reduced to a single GLENIM
\begin{equation}%
\begin{array}
[c]{c}%
i\frac{\partial E}{\partial z}+\sigma\frac{\partial^{2}E}{\partial t^{2}%
}+\sigma_{\perp}\Delta_{\perp}E+\kappa\left\vert E\right\vert ^{2}%
E+\nu\left\vert E\right\vert ^{4}E=\\
\\
i(\delta E+\beta\frac{\partial^{2}E}{\partial t^{2}}+\theta\Delta_{\perp
}E+\eta\left\vert E\right\vert ^{2}E+\gamma\left\vert E\right\vert ^{4}E)
\end{array}
\end{equation}
with the renormalisation ($2/\left\vert k^{\prime\prime}\right\vert
)^{1/2}t\rightarrow t$, $\Delta_{\perp}/2\left\vert n\right\vert
\varpi\rightarrow\Delta_{\perp}$, and $(\left\vert \mathbf{\chi}_{r}%
^{(3)}\right\vert \varpi/2)^{1/2}E\rightarrow E$.\ In reality, every material
exhibits losses. The dissipative parameter of linear loss $\delta=-\varpi
\Im\lbrack\varepsilon\mu]/2n$ is always negative insuring background stability
since in NIMs $\Im\lbrack\varepsilon\mu]<0$ \cite{Veselago}. The spectral
filtering term is positive taking into account that $\beta=\Im\lbrack
W]/(n\varpi\left\vert k^{\prime\prime}\right\vert )>0$ since $\Im\lbrack
W]<0$, as well as the energy diffusion term with $\theta=2k_{i}/k_{r}^{2}>0$
where $k_{i}>0$. Positive $\eta=\mathbf{\chi}_{i}^{(3)}/\left\vert
\mathbf{\chi}_{r}^{(3)}\right\vert $ corresponds to the cubic gain
compensating linear and nonlinear losses, thus, $\gamma=\mathbf{\chi}%
_{i}^{(5)}/\left\vert \mathbf{\chi}_{r}^{(3)}\right\vert $ must be negative
\cite{Skarka}. The parameter $\sigma=sgn(-k^{\prime\prime})$ is either
positive for anomalous GVD (AGVD) or negative for normal GVD (NGVD) (see
Fig.1). In NIMs the diffraction has a negative sign since the coefficient
$\sigma_{\perp}=sgn(n)<0$. Therefore,\ bright spatial solitons exist only if
the cubic nonlinearity is of the same sign as diffraction, thus,
$\kappa=sgn(\mathbf{\chi}_{r}^{(3)})<0$. Consequently, the quintic
nonlinearity has to be positive $\nu=\mathbf{\chi}_{r}^{(5)}/\left\vert
\mathbf{\chi}_{r}^{(3)}\right\vert >0$. In order to generate bright
spatiotemporal solitons so-called light bullets both diffraction and
dispersion must be compensated by saturating nonlinearity \cite{Skarka}.
Hence, both linear and nonlinear effects need to have the same sign.\ As a
consequence, light bullets can be generated only in NIMs with NGVD in the
range $0.7<\varpi<0.8$ for the choice of dissipative parameters $\beta$ and
$\delta$ as in Fig. 2. The lack of space limits our studies here to temporal
solitons described by temporal GLENIM (TGLENIM) corresponding to Eq. (4)
without spatial second derivatives. The generation, propagation, and stability
of light bullets in NIMs are investigated elsewhere.\ The obtained
nonintegrable TGLENIM can be solved only numerically. However, some analytical
approach even thought approximate is needed in order to have a better physical
inside. The approximate resolution of TGLENIM is done using variation method
extended to dissipative systems \cite{Skarka}. Although it is unable to
account structural change of pulse profile, this method can serve as a
guideline for simulation. A trial function $E=AA_{\ast}\exp\left[
t^{2}(iC-0.5T^{-2})T_{\ast}^{-2}+i\psi\right]  $ is chosen with amplitude
$A(z)$, pulse temporal width $T(z)$, wave front curvature $C(z)$, and phase
$\psi(z)$. Using renormalisation of the propagation variable $z/T_{\ast}%
^{2}\rightarrow z$ and linear loss parameter $\delta_{o}=T_{\ast}%
^{2}\left\vert \delta\right\vert <<1$, with $T_{\ast}=2^{1/2}\left(
4/3\right)  ^{3/4}$, and$\quad A_{\ast}=\left(  3/2\right)  ^{3/4}2^{-1/2}$,
these functions are optimized in order to yield corresponding Euler-Lagrange
ordinary differential equations%
\begin{equation}
\frac{dA}{dz}=(\delta_{o}+\frac{5\eta}{2}A^{2}+2\gamma A^{4}-\frac{\beta
}{T^{2}}-2\sigma C)A,
\end{equation}%
\begin{equation}
\frac{dT}{dz}=(4\sigma C-\eta A^{2}-\gamma A^{4})T+\frac{\beta}{T}-4\beta
T^{3}C^{2},
\end{equation}%
\begin{equation}
\frac{dC}{dz}=-4\sigma C^{2}+\frac{\sigma}{T^{4}}-\kappa\frac{A^{2}}{T^{2}%
}-\nu\frac{A^{4}}{T^{2}}-4\frac{\beta}{T^{2}}C,
\end{equation}%
\begin{equation}
\frac{d\psi}{dz}=-\frac{\sigma}{T^{2}}+\frac{5}{2}\kappa A^{2}+2\nu
A^{4}+2\beta C.
\end{equation}
Solving Eqs. (5-7) with zero $z$ derivatives,\ a double steady-state solution
with amplitudes $A_{+}$ and $A_{-}$ is obtained $A_{\pm}^{2}=\frac
{(\beta\kappa\sigma-4\eta)\mathbf{\pm}\sqrt{(\beta\kappa\sigma-4\eta
)^{2}+8\delta_{o}(3\gamma-\beta\nu\sigma)}}{2(\beta\nu\sigma-3\gamma)}$. The
family of solitons in conservative systems, reduces into either double or
simple solution for each set of dissipative parameters. A double solution
($A_{-}>A_{+}$) exists in the ($\eta$, $\gamma$) domain between the parabola
$(\beta\kappa\sigma-4\eta)^{2}+8\delta_{o}(3\gamma-\beta\nu\sigma)=0$ and
straight line $A_{-}=1$ for NGVD and AGVD in Fig.3. Above $A_{-}=1$ persists
only $A_{+}$. The ratio between dissipative parameters $\beta$ and $\delta$
(given in Fig.2) is kept same for NGVD and AGVD, in order to stress the
similarity of domains. The beam power $P=3\sqrt{\pi/2}A^{2}T$ is no more
conserved in dissipative\ systems \cite{Skarka}. However, the temporal width
$T=A^{-1}(\sigma\kappa+\sigma\nu A^{2})^{-1/2}$ and the power depend, up to
$\upsilon=\max\{\left\vert \beta\right\vert ,\left\vert \delta\right\vert
,\left\vert \eta\right\vert ,\left\vert \gamma\right\vert \}$, only on the
amplitude.\ The striking difference from the conservative systems is the
nonzero wave front curvature $C=-A_{o}^{2}\left(  \beta\kappa-\eta
\sigma+(\beta\nu-\gamma\sigma)A_{o}^{2}\right)  4^{-1}$. To be soliton a
steady state solution has to be stable. Our stability criterion based on the
method of Lyapunov's exponents has to be extended to NIMs in order to check
the stability of each steady-state. The solution of Eqs. (5-7) is stable if
and only if \ the real parts of the solutions $\lambda$ of the equation
$\left(  \lambda^{3}+\alpha_{1}\lambda^{2}+\alpha_{2}\lambda+\alpha
_{3}\right)  =0$ are all nonpositive \cite{Skarka}. The stability criterion
for TGLENIM is satisfied when coefficients $\alpha$ fulfill\ Hurwitz
conditions
\begin{equation}%
\begin{array}
[c]{c}%
\alpha_{2}=4\kappa A^{4}(\kappa+\nu A^{2})>0,\\
\alpha_{3}=2A^{6}(\kappa+\nu A^{2})^{2}[2A^{2}(\beta\nu\sigma-3\gamma
)+(\beta\kappa\sigma-4\eta)]>0,\\
\alpha_{4}=4(\kappa+\nu A^{2})[A^{8}(12\gamma\kappa-8\eta\nu)+A^{6}%
(13\eta\kappa-8\delta_{o}\nu)]>0,
\end{array}
\end{equation}
where $\alpha_{4}=\alpha_{1}\alpha_{2}-\alpha_{3}$ taking into account that
$\alpha_{1}=A^{4}(4\beta\nu\sigma-6\gamma)+A^{2}(\beta\kappa\sigma-4\eta)$.
The steady-state solution $A_{-}$ satisfies this stability criterion only
between the parabola and curve $s$ corresponding to $\alpha_{4}=0$ in Fig.3.
The solution $A_{+}$ is everywhere unstable. However, the stability of the
solution $A_{-}$ is only a prerequisite to obtain a soliton after a
self-organizing evolution. Indeed, an input pulse chosen in the established
stable domain corresponds to a point on the upper stable branch of
analytically obtained bifurcation curve $v$\ for power $P$ versus control
parameter $\gamma$ in Fig.4. The curve $v$ is only a good approximation of the
numerically obtained exact bifurcation curve $n$ composed of solitonic
attractors for different parameters.\ The lower unstable branch corresponds to
unstable solutions $A_{+}$. A stable steady state on the curve $v$ taken as
the input for numerical simulations evolves toward the\ soliton on the curve
$n$. The domain of stability is checked point by point in order to confirm by
numerical simulations that corresponding inputs always lead to a soliton.
Indeed, the numerically obtained domain of stability limited by the curve
$s_{n}$ in Fig.3 is even slightly larger than the analytical one. Numerical
simulations confirm that a soliton is propagating with, analytically
predicted, nonzero wave front curvature. As a consequence, the dispersion is
overcompensated by saturating nonlinearity leading to the collapse
\cite{Skarka}. However, the collapse is prevented by losses equilibrated in
turn by gain. Therefore, the self-organization of dissipationless solitons is
based on crosscompensation. During their propagation such solitons render the
medium effectively dissipationless.\ \ In order to check soliton robustness
its amplitude is increased $40\%$ at each $z$\ systematically $3000$ times
(see Fig.5). Such tremendous perturbations drastically increase the pulse
temporal width and amplitude. However, far from being annihilated the pulse,
remaining in attraction domain without losses and dispersion, maintains its
new form. Indeed, to each increase the self-organized system reacts lowering
its amplitude \cite{Skarka}. Perturbations arrested the soliton recovers its
initial shape after about $400$ steps in $z$ giving evidence of astonishing
robustness. Perturbations may correspond to monochromatic EM pulses injected
in such novel active medium composed of dissipationless solitons and NIM
(SOLINIM). Therefore, it seems that SOLINIM behaves for other pulses as
effectively dissipationless.

In conclusion, SOLINIM may be considered as a novel very promising active
composite medium due to the synergy between dissipationless solitons and NIM.
In order to describe such SOLINIM systems, newly established coupled
nonparaxial Helmholtz equations for electric and magnetic fields are reduced
to novel paraxial complex cubic-quintic Ginzburg-Landau equation. In contrast
with ordinary media, bright dissipationless light bullets may propagate only
in NGVD NIMs. However, bright dissipationless temporal solitons can be
generated in NIMs with both NGVD and AGVD. A stability criterion is
established. Choosing as input a steady-state with dissipative parameters from
obtained stability domains for either NGVD or AGVD NIMs, a self-organized
propagation always results in generation of extremely robust dissipationless
soliton. Such solitons during their propagation render NIM effectively
dissipationless. We hope that in SOLINIMs the practical realization of
peculiar effects like the cloaking and the superresolution will be no more prevented.

\textbf{ACKNOWLEDGMENTS}

This research has been in part supported by French--Serbian cooperation,
CNRS/MSCI agreement no. 20504. The work of VIB was supported by ISTC grant
G1366. Work at the Institute of Physics is supported by the Ministry of
Science of the Republic of Serbia, under the project OI 141031.\smallskip\

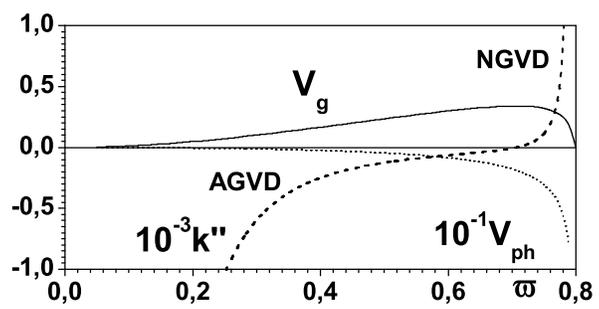
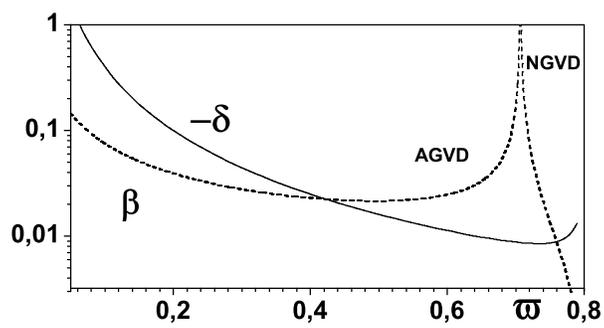
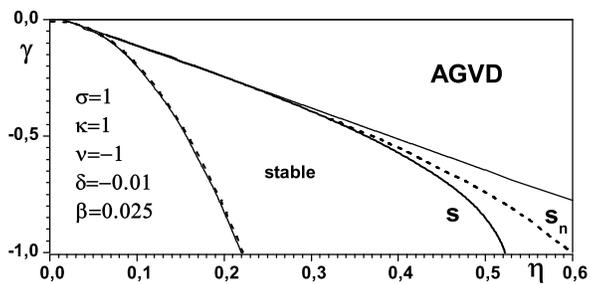
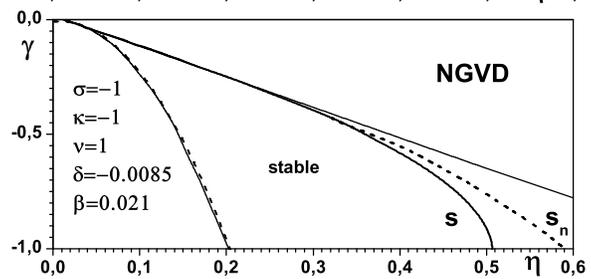
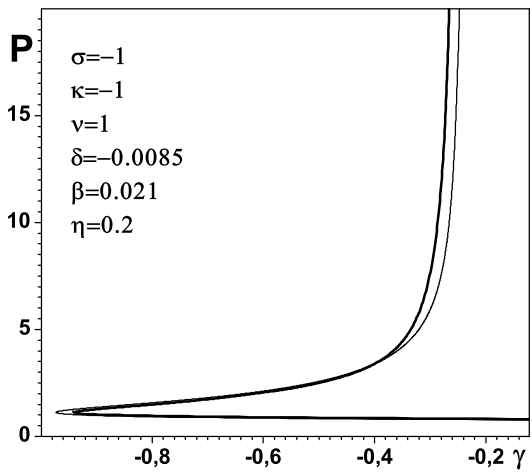
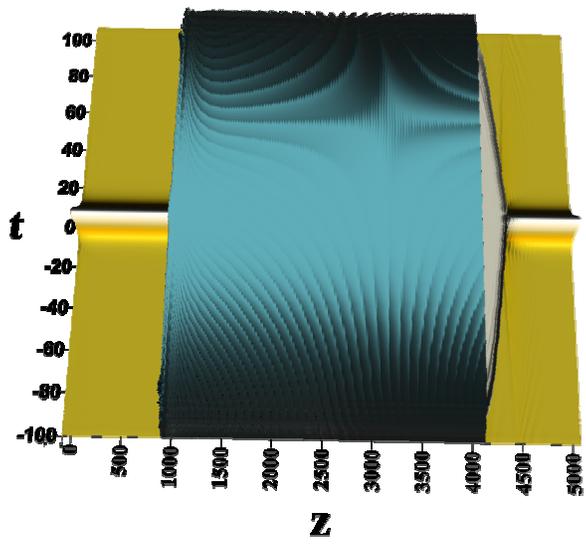
\end{document}